\documentclass[10pt,a4paper]{article}
\usepackage[english]{babel}
\usepackage{verbatim}
\usepackage{graphicx}
\usepackage{bm}
\usepackage{mathrsfs}
\usepackage[utf8x]{inputenc}
\usepackage{amsfonts}
\usepackage{amssymb}
\usepackage{amsmath}
\usepackage{hyperref}
\hypersetup{colorlinks=true}
\newcommand{\F}{F}
\newcommand{\vev}[1]{ \left\langle #1 \right\rangle}
\def\3F2{\mbox{$\,_3${F}$_2\!$}}
\def\2F1{\mbox{$\,_2${F}$_1\!$}}

\renewcommand{\vec}[1]{\mathbf{#1}}
\newcommand{\pb}{P_b}
\newcommand{\pc}{P_c}
\newcommand{\pcr}{\bar{P}_{c}}
\newcommand{\Lc}{L_c}
\newcommand{\Ham}{\mathscr H}
\newcommand{\s}{S}
\newcommand{\gam}{\Gamma}
\newcommand{\B}{B}
\DeclareMathOperator{\tw}{tw}
\DeclareMathOperator{\wrd}{wr}
\newcommand{\lk}{lk}

\DeclareMathOperator{\Lk}{Lk}
\DeclareMathOperator{\Tw}{T\!w}
\DeclareMathOperator{\Wr}{W\!r}
\DeclareMathOperator{\Det}{Det}
\newcommand{\free}{\mathfrak{f}}
\newcommand{\fth}{\mathfrak{f}_{th}}
\newcommand{\ftw}{\mathfrak{f}_{twist}}
\newcommand{\fh}{\mathfrak{f}_{pot}}
\newcommand{\fconf}{\mathfrak{f}_{conf}}
\newcommand{\de}{\mathrm{d}}

\begin{document}
\author{Marc Emanuel\\Theoretical Soft Matter and Biophysics, Institute of Complex
Systems, \\Forschungszentrum J\"ulich, J\"ulich 52425, Germany}
\title{On the confinement of semiflexible chains under torsion}
\maketitle

The following article has been accepted by JCP. After it is published, it will be found at  \href{http://link.aip.org/link/?jcp}{J. Chem. Phys.}. 
\section{Introduction}
The mechanical properties of biopolymers like DNA and actin are well
described by a model of semiflexible polymers, characterized by a
tangent-tangent correlation length that is considerably larger than the
monomer size. A proper mathematical description is the worm-like chain
(WLC): an elastic space curve with bending modulus $A$, that directly relates to this
correlation length or orientational persistence length $\pb=A/k_BT$.

In the living cell biopolymers are usually in a crowded, strongly confined
environment. 
Understanding the essential difference in statistics between the two
extremes of the mechanical, 
strongly confined limit and the critical point of the self
avoiding random walk limit, was an important development in polymer
physics. In the first limit, the so called Odijk regime, the
confining
potential dominates on length scales shorter than the persistence length while
in the latter case, the deGennes regime, the excluded volume interaction
dominates the thermal fluctuations on length scales of the channel diameter.
The general picture is the following~\cite{Odijk:2008}. Starting
from strong confinement in a narrow channel: the polymer has due to its
stiffness no other option than to follow the channel monotonously. The length
scale over which the polymer is unperturbed by the confining
potential, the so called deflection or collision length, is the dominating
length scale. Upon increasing the channel size, or decreasing the potential
strength, hairpins will form inverting the direction of the backbone of the
polymer. The average distance between hairpins, the so called global persistence
length, decreases with increasing channel size and becomes the length scale that
determines the span of the polymer until this global persistence length becomes
of the
order of the orientational persistence length. At that point the a transition
regime sets in in which the volume interactions, needed in the traditional de
Gennes regime, are non existent relative to the thermal energy. This is a
consequence of the slender shape of the polymer. Finally for channel sizes
considerably larger than the orientational persistence length, when the
channel-length
over which the volume interactions have to be added to account for at least $1\:
k_BT$ shrinks to the diameter of the channel, the de Gennes description becomes
applicable.

The reduction in degrees of freedom caused by the confinement, can be expressed
as an entropic cost that is
often a decisive ingredient to statistical physics calculations, for
systems in thermal equilibrium, but also for the dynamics of the
system.
Examples are the nematic transition of semiflexible polymers, the
ejection of DNA from viral capsids~\cite{Leforestier:2010}, translocation of
DNA through nanopores~\cite{Meller:2000},
the elongation and dynamics of DNA in nanochannels~\cite{Reisner:2004} and the
dynamics of actin
networks~\cite{Gardel:2004}.
From an application point of view the possibility to lay out DNA almost fully
stretched in a nanochannel makes it possible to study DNA and its interaction
with proteins on the base pair level~\cite{reisner2012dna}.

In many cases the WLC model is not sufficient though and it becomes crucial to
understand how twist effects change this picture. This is for example the case
in
understanding the physics of transcription, the dynamics of supercoiling of DNA
in vivo and the torque molecular motors can produce. In this paper the
influence of torque on the confinement of a semiflexible chain is studied in
the Odijk regime. It will be shown
how as a consequence fluctuations in the confined directions get coupled. Also
the influence of torque on the global persistence length is studied. An
important result is a dramatic decrease of the elongation of DNA in
nanochannels under moderate torques.

The paper is organized as follows: In section~\ref{sec:1} we set up the
model
and its Hamiltonian. We first consider an
isotropic
channel and show how the parametrization can be changed from a potential
strength
to a deflection length. We next proceed to treat the general anisotropic case.
In section~\ref{sec:2} we discuss some refinements, like the addition
of a stretch- and twist-stretch--moduli, how to
map the model to
hard-walled channels and we briefly discuss how torque influences the
elongation problem in nanochannels.

\section{The model}\label{sec:1}
The free energy of a freely rotating semiflexible chain in a
confining harmonic potential was calculated
in Ref.~\cite{Burkhardt:1995}.
These calculations we now perform for a persistent chain under torsion.
In fact confinement by a constant harmonic
potential is usually not what is needed. In case of a hard-walled nanochannel,
the width of the channel is better approximated by a fixed standard deviation
than a harmonic potential. The confining potentials in vivo and in single
molecule experiments are usually also far from harmonic,
while the standard deviation of the channel is used as a
variational parameter~\cite{Ubbink:1999,Emanuel:2012}.
Nonetheless the harmonic potential is used as a Lagrange multiplier, an
artificial confining potential to set the standard deviation $\sigma$ of the
transversal polymer distribution as measure of width of the confinement. The
benefit is an analytically solvable model and as long as the confinement is
strong enough specification of the second moment of the chain distribution will
do.

The typical length scale, that determines the
physical
properties is the deflection-length $\lambda=\sqrt[3]{\pb\sigma^2}$
of the chain, the length above which confinement dominates thermal
fluctuations.
The main calculations will be performed in a fixed number of turns, or linking
number
$\Lk$, ensemble. To keep the treatment transparent we model the persistent
chain
as a framed space-curve with a persistence-length $P_b$ and a torsional
persistence length $\pc$. The results are easily extended with a finite stretch
modulus $\s$ and a twist-stretch coupling, $\B$~\cite{Marko:1998}. All energies
are scaled by $k_BT$ (unless mentioned) for notational reasons. As a result
forces have the dimension of an inverse length.

The
chain is
confined in 
two perpendicular directions, say $x$ and $y$, by a harmonic potential. As
in Ref.~\cite{Burkhardt:1995}, we assume the confinement to be strong
enough
for the
$z$-coordinate to be a single valued function of the curve. The framing is
needed
to describe the twist degree of freedom. Since the confinement potential is
localized in space we cannot resort to the usual tangential description but have
to start from the space curve~\cite{Burkhardt:1995}. 
The Hamiltonian is:
\begin{equation}\label{eq:hamstart}
\begin{split}
\Ham ={}& \int_0^{\Lc}\! \de s \left[\frac{\pb}{2}\left(\frac{\de^2
\vec{r}(s)}{\de s^2}\right)^2\! +\frac{\pc}{2}
\dot{\psi}^2(s)\right.\\
&+\left.\frac{b_x}{2}r_x^2(s)+\frac{b_y}{2}r_y^2(s)\right
] .
\end{split}
\end{equation} 
As parametrization we choose the arc-length of the chain. The strength of the
 harmonic potential is set by $b_{x,y}$. The twist angle $\psi(s)$ is defined
as follows:
let $\vec{t}(s)=\dot{\vec{r}}(s)$ be the tangent and $\vec{n}(s)$ be one of the
unit normals that define the local frame. The twist density is defined as the
differential number of rotations of $\vec{n}$ around the tangential direction:
$\tw(s)=\frac{1}{2\pi}(\vec{t}\wedge\vec{n})\cdot\vec{\dot{n}}$, which is a
total differential and so we can define $\dot{\psi}(s):=2\pi\tw(s)$,

The twist density is not a small parameter and it is not independent of the
radial fluctuations. To perform a perturbation expansion
we split of the independent fluctuation degrees of freedom. This we can do
using White's relation~\cite{White:1969,Fuller:1971}, 
relating linking number, twist $\Tw$, and writhe $\Wr$ (the Gauss linking number
of a
closed curve with itself):
\begin{align}\label{eq:white}
 \Tw&:=\int_0^{L_c}\de s\tw=\Lk-\Wr
\end{align}
By assumption the chain is almost fully elongated permitting us to take periodic
boundary conditions. With this choice are the tangents at the ends of
the chain parallel. Since $2\pi$ times the writhe+1 of a closed curve is
equal to the surface on the direction sphere enclosed by the path of the tangent
mod $4\pi$~\cite{Fuller:1978}, and the tangents at both ends are
parallel, we can define the writhe of the chain as the area enclosed by the
tangent along the contour, implicitly setting the $0$ of the writhe to be the
$z$-axis.
Since the $z$-coordinate is single valued we can use
Fuller's equation~\cite{Fuller:1978,Aldinger:1995} turning the writhe of the
chain in an integral over a writhe density $\omega$ that is up to quadratic
order in the fluctuations:
\begin{align}\label{eq:writhe}
 \wrd(s)&\simeq
\frac{1}{4\pi}(\dot{r}_x(s)\ddot{r}_y(s)-\ddot{r}_x(s)\dot{r}_y(s))
\end{align}
The linking number density we define as $\lk:=\Lk/\Lc$. Now define a fluctuation
field $\phi(s):=\dot{\psi} -2\pi\lk-2\pi\wrd(s)$. From Eq.~\eqref{eq:white} it
follows that $\int_0^{\Lc}\de s \phi(s)=0$. We use this to eliminate the twist
term and can now integrate out the independent field $\phi$.
Making use of Eq.~\eqref{eq:writhe} results up to second order and an
irrelevant constant in:
\begin{equation}\label{eq:hamnonisotrop}
\begin{split}
\Ham={}& \int_0^{\Lc}\de
s\left[\frac{\pb}{2}\vec{\ddot{r}}^2\!+\frac{b_x}{2}r_x^2+\frac{b_y}{2}
r_y^2\right.\\
&-\left.\pi \pc\lk(\dot{r}_x\ddot{r}_y-\ddot{r}_x\dot{r}_y)
\right]+2\pi^2\pc\lk^2\Lc
\end{split}
\end{equation} 
The first term contains up to quadratic order only the $x$ and $y$ components
since the $z$ component of the tangent vector $\vec{t}(s)$ is quadratic in the
fluctuations:
\begin{align}
 \vec{\ddot{r}}^2(s)=\dot{t}_x(s)^2+\dot{t}_y(s)^2+\dot{t}_z(s)^2\simeq
\dot{t}_x(s)^2+\dot{t}_y(s)^2
\end{align}
In the following we will take $\vec{r}$ to be two-dimensional.
For reasons of clarity we present the analysis in more detail for an isotropic
confinement
\subsection{Isotropic confinement}
For isotropic confinement we have $b:=b_x=b_y$.  We can get rid of some of the
coupling constants by appropriately
scaling the contour lengths and the channel width~\cite{Burkhardt:1995}
\begin{align}
\hat{s} &= b^{1/4}\pb^{-1/4}s& \breve{r_i} &=
\pb^{1/8}b^{3/8}\vec{r_i}
\end{align}
and so $\hat{\pc}=b^{1/4}\pb^{-1/4}\pc$ but $\hat{\lk}=b^{-1/4}\pb^{1/4}\lk$.
This scaling results in:
\begin{equation}\label{eq:hamcompl}
\begin{split}\Ham={}& \int_0^{\hat{\Lc}} \de \hat{s} \left[\frac{1}{2}\ddot{
      \breve{\vec{r}}}^2(\hat{s})
+\frac{1}{2}\breve{\vec{r}}^2(\hat{s})-\frac{1}{2}\zeta(\dot{\breve{r}}_x\ddot{
\breve { r } } _y-\ddot{
\breve { r } } _x\dot { \breve { r } } _y)
\right]\\
 &+2\pi^2\hat{\pc}\hat{\lk}^2\hat{\Lc}
\end{split}
\end{equation}
with
\begin{align}
\zeta=\frac{2\pi\pc\lk}{\pb^{3/4}b^{1/4}}
\end{align}
The configurational partition sum is a path integral over the two-dimensional
field $\breve{\vec{r}}$:
\begin{align}
 Z=\int \mathcal{D}(\breve{\vec{r}})e^{-\Ham}
\end{align}
The last term of the Hamiltonian is constant and can be taken out from under the
path integral. It represents the unperturbed twist energy. What is left is the
fluctuation part which can be
diagonalized, after a Fourier mode expansion using periodic boundary conditions,
with eigenvalues:
\begin{align}
 ev_{n,\pm} &=p_n^4+1 \pm
\zeta p_n^3 &
\text{with }p_n&=\frac{2\pi n}{\hat{\Lc}}
\end{align}
Negative eigenvalues appear when $\zeta \geq \zeta_{cr}=4/3^{3/4}$
and so one can speak of a critical linking number density:
$\lk_{cr}=2b^{1/4}\pb^{3/4}/(3^{3/4}\pi\pc)$.
The partition sum $Z=\int [\mathcal{D}\vec{r}]\exp[-\Ham]$, becomes a product of
Gaussian integrals, that can be compared to the
free chain~\cite{Kleinert}. The
result can be written as:
\begin{align}\label{eq:partiso}
 Z&=\mathcal{N}\exp(-2\pi^2\hat{\pc}\hat{\lk}^2\hat{\Lc})\prod_{n=1}^{\infty
}\frac{p_n^8}{ev_{n,+}ev_{n.-}}
\end{align}
with $\mathcal{N}$ collecting constant terms.
In the denominator of the product we have a quartic expression in $n^2$:
Let
$\eta_i,\bar{\eta}_i (i=1,2)$ be the pairs of conjugate roots:
\begin{align}
 (\eta_i^2+1)^2-\zeta^2\eta_i^3=0.
\end{align}
Then we can
write
the
product as:
\begin{align}
 \prod_{n=1}^{\infty}(\cdots)&=\prod_{i=1,2}\prod_{n=1}^{\infty}\frac{1}{
\left(1-\frac{\eta_i}{n^2}\right)}\frac{1}{\left(1-\frac{\bar{\eta}_i}{n^2}
\right)}
\nonumber \\
&=
\prod_{i=1,2}\frac{\pi\sqrt{\eta_i}}{\sin(\pi\sqrt{\eta_i})}\frac{\pi\sqrt{\bar{
\eta}_i}}{\sin(\pi\sqrt{\bar{\eta}_i})}
\end{align}
The complex part of the square root of the $\eta_i$'s give an exponential
decay with increasing chain length. 
Now write $\sqrt{\eta_{1,2}}=\frac{\hat{L}}{2\pi}(x_{1,2}\pm\imath y_{1,2})$. In
the long chain limit, as long as the $y_i$ are non-zero we obtain the free
energy density as:
\begin{align}
\free =\lim_{\Lc\to\infty} -\frac{1}{\Lc}\log Z &=
2\pi^2\pc\lk^2 +\frac{\bigl(\pb^3b\bigr)^{1/4}} {\pb}(\vert y_1\vert+\vert y_2
\vert)
\end{align}
\begin{figure}[htb]
  \includegraphics[width=\columnwidth]{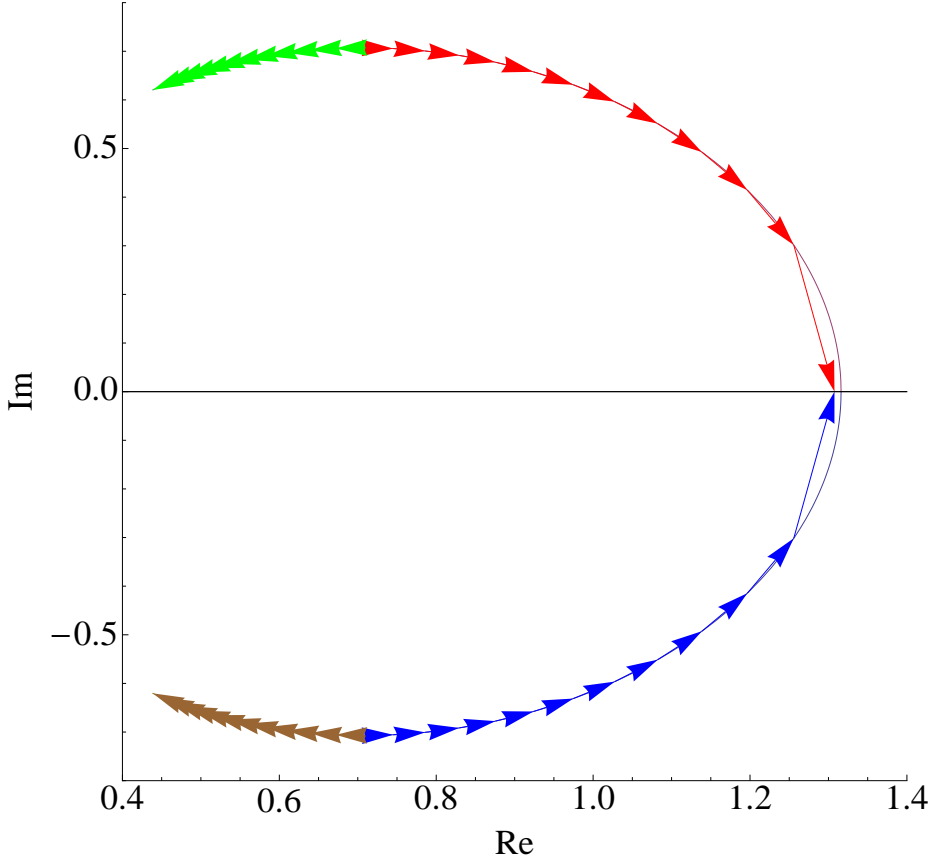}
  \caption[Flow of the roots $\sqrt{\eta_i}$ in the complex plane.]{Flow of the
roots $\sqrt{\eta_i}$ in the complex plane under increasing torsion. The
size of the arrows indicates the speed with which the roots flow as
explained in
the text.}
  \label{fig:rootflow}
\end{figure}
The first term is the zero-temperature twist energy density stored in the chain,
the second $\fth$
the thermal fluctuation contribution in presence of the confining
potential. In case the linking number
density
is zero, the roots are up to a sign equal: $
\vert y_{1,2}\vert =1/\sqrt{2}$ and we retrieve 
Burkhardt's result~\cite{Burkhardt:1995}. The flow of the roots under increasing
linking number density is depicted in Fig.~\ref{fig:rootflow}. Starting from
$\lk=0$ the root pairs move apart. Both decrease their imaginary parts,
decreasing the free energy.  This does not happen
in a symmetric way, as is shown by the line segments along trajectories,
where each segment
corresponds to a fixed increase of $\lk$, the arrow pointing in the direction
of the flow. At $\lk_{cr}$ the right pair of roots
becomes real indicating a singularity in the partition sum. 

As explained above the harmonic potential is only there to set the confinement.
The target is to express all quantities in the standard deviations of the
transversal distributions of the chain in the channel. To do this
and to extend the calculations to a non-isotropic confinement we perform the
previous calculations in an alternative way, taking first the logarithm
of Eq.~\eqref{eq:partiso} and then the infinite chain limit:
\begin{align}\label{eq:freecont}
 \fth={}&-\frac{1}{\Lc}\log Z\notag\\
={}&
-\frac{1}{\Lc}\sum_{n=1}^{\infty}\log\left(\frac{p_n^8}{(p_n^4+1)^2-\zeta^2p_n^6}
\right)+\mathcal{O}\big(\tfrac{\log(\Lc)}{\Lc}\big)\nonumber \\
\simeq{}& -\frac{\bigl(\pb^3b\bigr)^{1/4}}{\pb}\int_{0}^{\infty}\frac{\de
p}{2\pi}\log\left(\frac{p^8}{(p^4+1)^2-\zeta^2p^6}\right)
\nonumber \\
={}&-\frac{2\pi\pc\lk}{\pb\zeta}\left[-\sqrt{2}-\int_{0}^{\infty
} \frac { \de
p}{2\pi}\log\left(1-\frac{\zeta^2p^6}{(p^4+1)^2}\right)\right]
\end{align}
expanding the logarithm, making use of the fact that $\zeta p^3 < p^4+1$, for
$\zeta<\zeta_{cr}$, and noting that the series converges uniformly in any closed
interval contained in $[0,\zeta_{cr})$ this results in:
\begin{align}
\fth={}&-\frac{2\pi\pc\lk}{\pb\zeta}\left[-\sqrt{2}+\sum_{n=1}^{\infty}
\frac{1}{n}\int_{0}^{\infty}\frac{\de
p}{2\pi}\left(\frac{\zeta^2p^6}{(p^4+1)^2}\right)^n\right]\notag\\
={}&\frac{2\pi\pc\lk}{\pb\zeta}\sum_{n=0}^{\infty}
a_n\zeta^{2n}\qquad\text{with: } a_n:=-
\frac{\gam(\frac{n}{2}-\frac{1}{4})\gam(\frac{3n}{2}+\frac{1}{4})}{4\pi(2n)!}\\
\intertext{This can be written as a linear combination of  generalized hypergeometric functions:}
\fth={}&-\frac{2\pi\pc\lk}{\pb\zeta}\sqrt{2}\left[1-\3F2\left(-\tfrac{1}{4},\tfrac{1}{12},\tfrac{5}{12};\tfrac{1
}{4},\tfrac{1}{2};\tfrac{\zeta^4}{\zeta_{cr}^4}\right)\right.\notag\\
&\left.
\qquad+\frac{3\zeta^2}{32}\3F2\left(\tfrac{1}{4},\tfrac{7}{12},\tfrac{11}{12}
;\tfrac { 3
}{4},\tfrac{3}{2};\tfrac{\zeta^4}{\zeta_{cr}^4}\right)\right],\notag
\end{align}
making the bifurcation point $\zeta_{cr}$ explicit.
Defining:
\begin{align}
x&:=\frac{\zeta}{\zeta_{cr}}&\gamma&:=\sqrt{2}\frac{2\pi\pc\lk}{\pb\zeta_{cr}},
\end{align}
results in a free energy
density given by:
\begin{align}
\frac{\fth}{\gamma}&=\sum_{n=0}^{\infty}a_n
x^{2n-1}=\frac{1}{x}-\frac{x}{2\sqrt{3}}-\frac{5x^3}{72}+\mathcal{O}
(x^7)
\end{align}
From the series expansion of $\fth$ it is straightforward to find a series
expansion for the standard deviation $\sigma_r=\sqrt{\vev{\vec{r}^2}/2}$ of the
transversal position of the chain in the channel:
\begin{align}
\sigma_r^2=\frac{\partial
\fth}{\partial b}=-
\frac{1}{\pb\gamma^3}\sum_{n=0}^{\infty}(2n-1)a_n
x^{2n+3}, \label{eq:z_series}
\end{align}
which is again a linear combination of  generalized hypergeometric functions. We now
apply Lagrange inversion to this series in order to eliminate $b$ in favor of
$\sigma_r$ as dependent parameter for the free energy density.
To be able to use Lagrange inversion
the derivative at the point of inversion should be non-zero, which requires
non-zero coefficients of
the linear term. The
lowest term in Eq.~\eqref{eq:z_series} is cubic, so if we take the cubic root of
this series we obtain a series to which one
can apply Lagrange inversion. Introducing the deflection length,
$\lambda:=(\pb\sigma_r^2)^{1/3}$ we start from
its series expansion:
\begin{align}
\begin{split}
\lambda(x)={}&
\frac{\sqrt[3]{a_0 }}{\gamma}x\\
&+\frac{\sqrt[3]{a_0}}{\gamma}x\sum_{k=1}^{\infty}
\frac{\gam(-\tfrac{1}{3}+k)}{k!\gam(-\tfrac{1}{3})}\left(\sum_{n=1}^{\infty}
(2n-1)\frac{a_n } { a_0 } x^ {
2n}\right)^k,\label{eq:xtolambda}
\end{split}
\end{align}
and apply Lagrange inversion to this:
\begin{align}
 x(\lambda)&=\sum_{n=1}^{\infty}\frac{1}{n!}\left.\frac{\de^{n-1}}{\de
x^{n-1}}\left(\frac{x}{\lambda(x)}\right)^n\right\rvert_{x=0}\lambda^n
\end{align}
The first terms of the resulting series are:
\begin{align}
x(\lambda)&=\gamma
\lambda-\frac{\gamma^3\lambda^3}{6\sqrt{3}}-\frac{7\gamma^5\lambda^5}{216}
+\mathcal {O} \left((\gamma\lambda)^7 \right)
\nonumber \\ 
\fth(\lambda)&=
\frac{1}{\lambda}-\frac{\gamma^2\lambda}{3\sqrt{3}}+\frac{5\gamma^6\lambda^5}
{243\sqrt{3}}
+\mathcal{O}({z^{7/3}}) \notag\\
\intertext{or in terms of the unscaled variables:}
\begin{split}
\fth={}&
\frac{1}{\lambda}\left[1-\frac{1}{2}\left(\pi\pc\lk\frac{\lambda}{\pb}
\right)^2\right.\\
&\left.+\frac{5}{24}\left(\pi\pc\lk\frac{\lambda}{\pb}\right)^6
+\mathcal{O}\Bigl(\left(\pi\pc\lk\frac{\lambda}{\pb}
\right)^8\Bigr)\right ] \label
{eq:fth}
\end{split}
\end{align}
This free energy density contains the enthalpic free energy of the harmonic
potential. But the harmonic potential is just a Lagrange multiplier to set the
standard deviation, thus this enthalpic part
of the confining potential should be subtracted in order to obtain the net
cost for confining the molecule. The enthalpic contribution is easily
calculated from the inverted series:
\begin{align}
 \fh={}&\frac{1}{2}b\vev{\vec{r}^2}=\frac{27}{16}\left(\frac{\pi\pc\lk}{\pb}
\right)^4\frac{\lambda^3}{x^4(\lambda)}\notag\\
\begin{split}
={}&\frac{1}{\lambda}\left[\frac{1}{4}+\frac{1}{4}\left(\pi\pc\lk\frac{\lambda}{
\pb }
\right)^2
+\frac{3}{8}\left(\pi\pc\lk\frac{\lambda}{\pb}\right)^4\right.\\
&\qquad\left.+\mathcal{O}\Bigl(\left(\pi\pc\lk\frac{\lambda}{\pb}
\right)^6\Bigr)\right ]
\end{split}
\end{align}
Subtracting this free energy density from Eq.~\eqref{eq:fth} gives the
confinement free
energy
density:
\begin{align}
\begin{split}
 \fconf={}&\frac{1}{\lambda}\left[\frac{3}{4}-\frac{3}{4}\left(\pi\pc\lk\frac{
\lambda}{
\pb }
\right)^2
-\frac{3}{8}\left(\pi\pc\lk\frac{\lambda}{\pb}\right)^4\right.\\
&\qquad\left.+\mathcal{O}\Bigl(\left(\pi\pc\lk\frac{\lambda}{\pb}
\right)^6\Bigr)\right ]
\end{split}
\end{align}
Note that the free energy of confinement seems to decrease with increasing
linking number, while the standard deviation is kept constant. This is actually
the twist energy reduction due to thermal writhe. The unperturbed twist energy
density was the last term in Eq.~\ref{eq:hamcompl} adding the perturbation
contribution we recover the effective twist free energy density:
\begin{align}
 \ftw&=2\pi^2\pc\left(1-3\frac{\pc\lambda}{8\pb^2}\right)\lk^2:=2\pi^2\pcr\lk^2,
\end{align}
where a ``renormalized'' torsional persistence length $\pcr$, is defined that
is of the same form as that of a WLC under
tension~\cite{Moroz:1998}.
From Eq.~\eqref{eq:hamnonisotrop} we see that the linking number density is
conjugate to the thermal writhe density in $\fth$, allowing us to calculate its
expectation value and from White's equation the expectation value of the twist
density:
\begin{align}
\vev{\wrd}&\simeq-\frac{1}{4\pi^2\pc}\frac{\partial \fth(lk,b)}{\partial
lk}=\frac{3\pc\lambda\lk}{8\pb^2}\\
\vev{\tw}&\simeq\lk-\vev{\wrd}=\left(1-\frac{3\pc\lambda}{8\pb^2}\right)\lk
\end{align}
Thermal motion decreases the extension of the chain. It is calculated by
adding a tension term in the channel direction to the Hamiltonian:
\begin{align}\label{eq:ficforce}
 \Ham\rightarrow \Ham + \F z&\simeq\Ham +
\int_0^{\Lc}\de s\F (1-\frac{1}{2}\dot{\vec{r}}^2(s)).
\end{align}
The extension in the $z$-direction is $\rho:=\Delta z/\Lc
=\partial_{\F}\free(\F)\vert_{\F=0}$.
After a mode expansion the fluctuation part can be diagonalized again with $\F$
dependent eigenvalues. The calculation of the free energy density goes as
before:
\begin{multline}\label{eq:isofree}
 \fth= -\frac{\bigl(\pb^3b\bigr)^{1/4}}{\pb}\left[\int_{0}^{\infty}\frac{\de
p}{2\pi}\log\left(\frac{p^8}{(p^4+\tfrac{\F}{\sqrt{\pb
b}}p^2+1)^2}\right)^{\mbox{}}\right.\\
\left.+\sum_{n=1}^{\infty}\frac{1}{n}\int_0^{\infty}\frac{\de
p}{2\pi}\left(\frac{\zeta^2p^6}{(p^4+\tfrac{\F}{\sqrt{\pb
b}}p^2+1)^2}\right)^n\right]
\end{multline}
For the contraction factor we obtain in terms of the deflection length:
\begin{align}\label{eq:isorho}
 \rho=1-\frac{\lambda}{2\pb}\left[1+\left(\pi\pc\lk\frac{\lambda}
{\pb}\right)^2+\mathcal{O}\Bigl(\left(\pi\pc\lk\frac{\lambda}{\pb}
\right)^4\Bigr)\right]
\end{align}
A last quantity that is easily extracted is the expectation value of the
torque $\tau$ one has to apply to keep the linking number fixed:
\begin{align}\label{eq:tau}
 \vev{\tau}=\frac{1}{2\pi}\frac{
\partial\ftw}{\partial\lk}=2\pi\pcr\lk
\end{align}
The influence of torque on the confined polymer is small for a strong
confinement. The restricted space around the polymer hinders the production of
thermal writhe. Upon increasing the standard deviation thermal withe becomes
stronger. In the
torsionless case, the demand that the z-coordinate is a single valued function
along the backbone results in the demand that the deflection length
is much smaller than the persistence length. With torsion, this upper bound
decreases with increasing torque. From the
series~\eqref{eq:isofree},~\eqref{eq:isorho} it can be seen that the demand on
deflection length changes to
\begin{align}
 \lambda\ll\frac{\pb}{\pi\pc\lk}.
\end{align}
Using Eq.~\eqref{eq:tau} this can be written as a torque dependent constraint
on the standard deviation:
\begin{align}
 \sigma\ll \pb \left(\frac{8\pb}{3\pc+4\pb \tau}\right)^{3/2}
\end{align}
Note that $\tau$ is scaled by $k_BT$. In
section~\ref{sec:2} the effect of this approaching instability on the
extension is discussed.
\begin{figure}[htb]
 \includegraphics[width=\columnwidth]{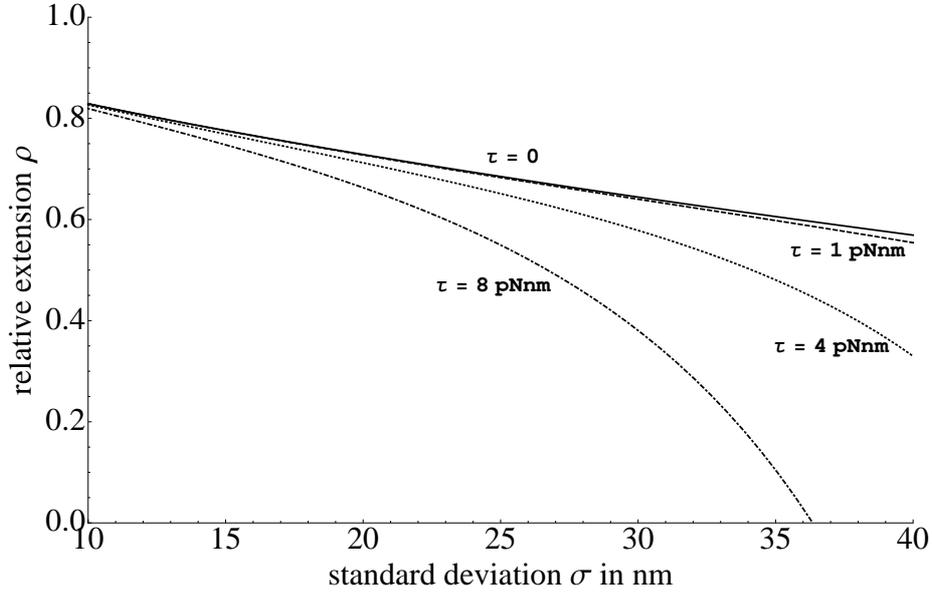}
  \caption{\label{fig:rhosigiso}The local relative extension for DNA in a
confined environment,
producing a standard deviation up to $8$ nm with a range of torques applied
along the backbone. The persistence lengths were chosen to be $\pb=50$nm and
$\pc=100$nm}
\end{figure}
In Figure~\ref{fig:rhosigiso} the effect of a finite torque is shown on the
semi-local extension of DNA, with torques that are realistic for torque
producing molecular motors in vivo\footnote{see section~\ref{sec:2}}. It is
important to realize that the
strength of the harmonic potential needed to obtain a fixed standard deviation
increases with increasing torque. 

\subsection{Non-isotropic confinement}
The strategy for the non-isotropic case will be the same as for the isotropic
case. The increased complexity is a consequence of the twist mixing fluctuations
in the two transversal directions.
The Hamiltonian is again block diagonalized after a Fourier expansion. With
\begin{align}
 b&:=\frac{33(b_x^2b_y+b_xb_y^2)+(b_x^2+14b_xb_y+b_y^2)^{3/2}
-b_x^3-b_y^3}{128b_xb_y}.
\end{align}
and a scaling like in the isotropic case we see that the expression for the
critical
point is, like in the isotropic case: $\zeta_{cr}=4/3^{3/4}$. We assume
$\zeta$ to be small. Defining $\bar{b}_{x,y}:=b_{x,y}/b$ the fluctuation
part of the free energy density can again be written as an integral over the
determinants:
\begin{multline}
\fth= -\frac{\bigl(\pb^3b\bigr)^{1/4}}{\pb}\left[\int_0^{\infty}\frac{\de
p}{2\pi}\log\left(\frac{p^8}{(p^4+\bar{b}_x)(p^4+\bar{b}_y)}\right)\right.\\
+\left.\sum_{n=1}^{\infty}\frac{\zeta^{2n}}{n}\int_0^{\infty}\frac{\de
p}{2\pi}\left(\frac{p^6}{(p^4+\bar{b}_x)(p^4+\bar{b}_y)}\right)^n\right]
\end{multline}
Assume $b_x\leq b_y$. Integration results in a power series in
$\zeta$, with coefficients that can be expressed
in series over hypergeometric functions. We will only need the first terms.
They are:
\begin{align}
\begin{split}
 \fth={}&\frac{\bigl(\pb^3b\bigr)^{1/4}}{\pb}
\left(\frac{\bar{b}_x^{1/4}+\bar{b}_y^{1/4}}{
\sqrt{2}}\right.\\&\left.-\frac{\bar{b}_x^{1/2}+\bar{b}_x^{1/4}\bar{b}_y^{1/4}
+\bar
{b}_y^{1/2}}{4\sqrt{2}(\bar{b}_x^{1/4}+\bar{b}_y^{1/4})(\bar{b}_x^{1/2}+\bar{b}
_y^{1/2})}\zeta^2+\mathcal{O}(\zeta^4)\right).
\end{split}
\end{align}
Since the terms in the series expansion are independent of the $b-$scale, we can
choose it as
$b=1/\pb^3$ and change from $b_{x,y}$ to the dimensionless $z_{x,y}:=\bigl(\pb^3 b_{x,y}\bigr)^{-1/4}$ resulting in
\begin{align}
\fth&=
\frac{1}{\pb}\left(\frac{z_x+z_y}{\sqrt{2}z_xz_y}-\frac{
z_xz_y(z_x^2+z_xz_y+z_y^2)}{4\sqrt{2}(z_x+z_y)(z_x^2+z_y^2)}
\zeta_1^2+\cdots\right),
\end{align}
with $\zeta_1=2\pi\pc\lk$.
To get an expression for the free energy as function of the deflection lengths
we will use a multivariate extension of Lagrange
inversion~\cite{Good:1960,Gessel:1987}.
As in the isotropic case we extract the standard
deviations for the two channel directions from the free energy density:
\begin{align}
 \sigma_x^2&=-\frac{\pb^3}{2}z_x^5\partial_{z_x}\fth\notag\\
&=\frac{\pb^2z_x^3}{2\sqrt{2}}\left(1+\frac{z_x^2z_y^4(3z_x^2+2z_xz_y+z_y^2)}{
4(z_x+z_y)^2(z_x^2+z_y^2)^2}\zeta_1^2+\cdots\right),
\end{align}
and $z_x\leftrightarrow z_y$ for $\sigma_y$. With
$f_{x,y}:=z_{x,y}/\lambda_{x,y}(zx,zy)$, we have a relation between
$\lambda_{x,y}$ and $z_{x,y}$ of the right type to apply Good-Lagrange
inversion, that is:
\begin{align}
 z_{x,y}&=f_{x,y}\lambda_{x,y} &\text{and } f_{x,y}(0,0)\neq 0.
\end{align}
Under these condition given any (formal) power series $g$ in $z_x$ and $z_y$ we
can express $g$ uniquely as a power series in $\lambda_x$, $\lambda_y$ by the
Good-Lagrange formula~\cite{Good:1960,Gessel:1987}:
\begin{align}\label{eq:goodlagrange}
  [\lambda_{x,y}&]^{n,m}g\bigl(z_{x}(u_x),z_{y}(u_{y})\bigr)\notag\\
={}&[z_{x,y}]^{n,m}g(z_x,z_y)f_x^n(z_x,z_y)f_y^m(z_x,z_y)\notag\\
&\qquad\Det\biggl(\delta_{i,j}-\frac{z_i}{f_j(z_x,z_y)}\frac{\partial
f_j(z_x,z_y)}{\partial z_i}\biggr)\nonumber \\
={}&
[z_{x,y}]^{n,m}g(z_x,z_y)f_x^n(z_x,z_y)f_y^m(z_x,z_y)\notag\\
&\qquad\frac{1}{9}
\Det\biggl(\frac{z_i}{\sigma^2_j(z_x,z_y)}\frac{\partial
\sigma^2_j(z_x,z_y)}{\partial z_i}\biggr),
\end{align}
where $[x_{1,2}]^{n,m}F(x_{1,2})$ denotes the coefficient of the monomial
$x_1^nx_2^m$ in the (formal) power series $F$. When applied to $\fth$ we obtain
a series that is symmetric against interchange of $\lambda_x$ and $\lambda_y$.
We define the symmetrized monomial:
\begin{align}
 [n,m]:=\lambda_x^n\lambda_y^m+\lambda_y^n\lambda_x^m.
\end{align}
For $\fth$ we get:
\begin{align}
 \fth={}&\frac{1}{2}[-1,0]
-\frac{\zeta_1^2
}{6\pb^2}([1,0]-[4,-3]\notag\\&\quad+[5,-4]-[8,-7]+\cdots)+\mathcal{O}
\Bigl(\zeta_1^6\Bigr)\notag\\
\simeq{}&\frac{1}{2}[-1,0]-\frac{\zeta_1^2}{6\pb^2}\frac{[3,1]+[2,2]/2}{[1,0][2,
0 ] }
\notag\\
={}&\frac{1}{2}(\lambda_x^{-1}+\lambda_y^{-1})-\frac{2(\pi\pc\lk)^2}{3\pb^2}
\frac {
\lambda_x^{3}\lambda_y+\lambda_x^{2}\lambda_y^{2}+\lambda_x\lambda_y^{3}}{
(\lambda_x+\lambda_y)(\lambda_x^2+\lambda_y^2)}
\end{align}
The combination of powers in the deflection lengths in the 2nd term will appear
several times. For this reason we define an effective twist deflection length
\begin{align}
 \bar{\lambda}&:=2\frac{
\lambda_x^{3}\lambda_y+\lambda_x^{2}\lambda_y^{2}+\lambda_x\lambda_y^{3}}{
(\lambda_x+\lambda_y)(\lambda_x^2+\lambda_y^2)}.
\end{align}
The factor $2$ is a matter of convention, chosen for comparison with the
$\pc$ renormalization concept of a chain under tension of
Ref.~\cite{Moroz:1998}.
Let $\lambda_x$ be the smaller of the two deflection lengths, corresponding to
the stronger confined direction, this effective deflection length has a value
in between $\frac{3}{2}\lambda_x$ for $\lambda_x=\lambda_y$ and
$2[1-(\lambda_x/\pb)^3]\lambda_x$ for $\lambda_x<<\lambda_y$. 

Apparently the thermal writhe effect on the free energy is dominated by the
smallest deflection length. This is understandable since writhe is non-planar
and since the polymer is directed along one axes, tightening one of the
transversal directions forces the polymer to be more planar. Just
like in
the isotropic
case we subtract the free energy contribution of the harmonic potential, to get
the pure confinement free energy. The calculations are merely a
repetition of the $\fth$ calculation so we just give the result up to the same
order:
\begin{align}
 \fh&=\frac{1}{8}(\lambda_x^{-1}+\lambda_y^{-1})+\frac{(\pi\pc\lk)^2}{6\pb^2}
\bar{
\lambda}\notag\\
\fconf&=\fth-\fh=\frac{3}{8}(\lambda_x^{-1}+\lambda_y^{-1})-\frac{(\pi\pc\lk)^2}
{2\pb^2}\bar{\lambda}.\label{eq:fconfaniso}
\end{align}
The second term can again be interpreted as a reduction of the twist free energy
density. Adding it to the unperturbed twist free energy density  we find:
\begin{align}
 \ftw&=2\pi^2\pc\left(1-\frac{\pc\bar{\lambda}}{4\pb^2}\right)\lk^2
=:2\pi^2\pcr\lk^2,
\end{align}
the ``renormalized'' torsional persistence length $\pcr$, is again of the
same form as the thermal softening of a WLC under
tension~\cite{Moroz:1998}.
The average writhe and twist densities follow from the same derivation as in the
isotropic case, and using the Good-Lagrange
formula for parameter change results in:
\begin{align}
 \vev{\wrd}=&\frac{\bar{\lambda}\pc\lk}{4\pb^2} &
\vev{\tw}=&\left(1-\frac{\bar{\lambda}\pc}{4\pb^2}\right)\lk
\end{align}
The extension is calculated like before by adding a force
term~\eqref{eq:ficforce}.
We perform the derivation by $\F$ before the
integration, choose $b=\pb^3$ and expand in $\zeta_1$. The resulting extension
is expressed in $z$
\begin{align}
\rho&=1-\frac{z_x+z_y}{4\sqrt{2}}-\frac{5}{16\sqrt{2}}\frac{z_x^3z_y^3}{
(z_x+z_y)(z_x^2+z_y^2)}\zeta_1^2-\mathcal{O}(\zeta_1^4)
\end{align}
Finally resorting again to Good-Lagrange inversion~\eqref{eq:goodlagrange} we
obtain
\begin{align}
\begin{split}
\rho={}&1-\frac{1}{4}\left(\frac{\lambda_x}{\pb}+\frac{
\lambda_y}{\pb}\right)\\&\quad-\frac{
(7\lambda_x^3\lambda_y^5-2\lambda_x^4\lambda_y^4+7\lambda_x^5\lambda_y^3)(\pi
\pc\lk)^2}{12(\lambda_x+\lambda_y)(\lambda_x^2+\lambda_y^2)^2\pb^3}.
\label{eq:shortaniso}
\end{split}
\end{align}
Also this last expression has the feature that it is the tightest direction that
determines the thermal writhe.
\begin{figure}[htb]
  \includegraphics[width=\columnwidth]{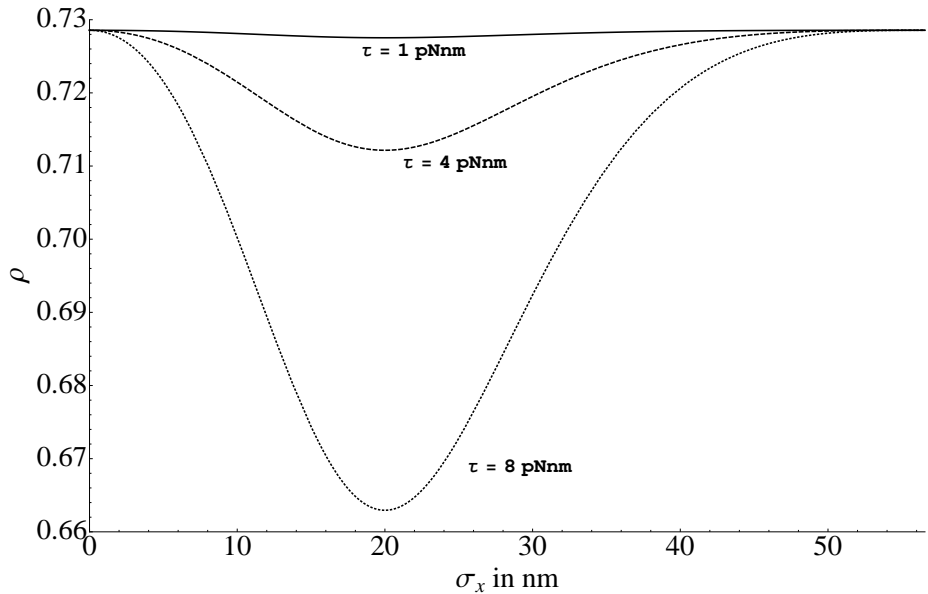}
  \caption{\label{fig:squeeze}The squeezing out of thermal writhe for DNA by
anisotropy for $3$ different torques while keeping the influence of bending
modes contant. The elastic moduli are chosen as in Fig.~\ref{fig:rhosigiso}}
\end{figure}
In Figure~\ref{fig:squeeze} the extension of DNA upon
non-isotropic confinement is shown. The amount of
confinement anisotropy is changed by varying the standard deviations such that
the sum of the two deflection lengths remains constant. In that way the
contraction due to bending degrees of freedom does not change. The isotropic
point is at $\sigma_x=\sigma_y=20$nm. In the limit of vanishing standard
deviation in one direction, thermal writhe is completely suppressed.

\section{Some extensions}\label{sec:2}
We did not include a twist-stretch coupling, that is typically nonzero  for
chiral
molecules,
or or a stretch modulus in the calculations. To include them
complicates the calculations only modestly. Mainly because the
stretch modulus is large compared to tensions normally applied, their
influence is small. In lowest order it amounts to a correction of the bare
torsional persistence length to $\pc':=\pc-B^2/S$, where $B$ is the
(dimensionless) twist-stretch coupling, and $S$ the stretch
modulus~\cite{Marko:1998}. Furthermore the contraction factor
acquires a
contribution: $\rho\rightarrow\rho-2\pi\B\lk/S$.

The calculations were performed using a harmonic potential as a substitute. To
relate this to a hardwalled channel we can extend the reasoning
from Ref.~\cite{Ubbink:1999} and calculate the deflection length of the
hard-walled
channel, with channel width $d$, by comparing the confinement free energy at
zero linking number density with the
numerical results from Ref.~\cite{Burkhardt:1995,Yang:2007}. This leads
to:
\begin{align}\label{eq:hard}
 \lambda_{\mathrm{hard}}=\frac{3}{8\times1.1036}(\pb d^2)^{1/3}
\end{align}
One can reason now that since we have the free energy as function of the
standard deviations, and thus up to this numerical factor of the channel
dimensions, we can extend this to a finite torque. Had we used the potential
strength b as parameter the chain would fluctuate through the channel wall.

A WLC under tension undergoes upon increasing linking number a buckling
transition to a plectoneme, a critical multi-plectoneme phase or to a multitude
of irregular configurations of loop-like structures (which might lead to
structural changes in the WLC)~\cite{Emanuel:2012}. Into which of these 3
categories the transition belongs depends on material and environmental
parameters, but the linking number density is in all cases controlled by a
mechanical instability at a critical linking number density of
$\lk_{cr}=\sqrt{\F\pb}/(\pi\pc)$. Comparing this with the instability in the
confined chain, we see that the deflection lengths also here have a comparable
role. The critical linking number density of the confined chain corresponds to
that
of a chain under tension $\F$ when:
$$\sigma\cong \left(\frac{\pb}{\F^3}\right)^{1/4}$$
As an important example we mention DNA where the maximum linking number density
causing a buckling transition is for forces of around $4$ pN
($\F=1\:\text{nm}^{-1}$), higher
forces causing a structural transition within the DNA double helix. With a
persistence
length of $50$ nm this would correspond to a channel dimension $\sim 3\!$ nm,
which is an order of magnitude smaller than the channels used in confinement
experiments. In a plectoneme on the other hand the fluctuations of the strands
can easily
have comparable standard deviations~\cite{Emanuel:2012}, showing that
plectonemes are indeed stable. 
\begin{figure}[htb]
 \includegraphics[width=\columnwidth]{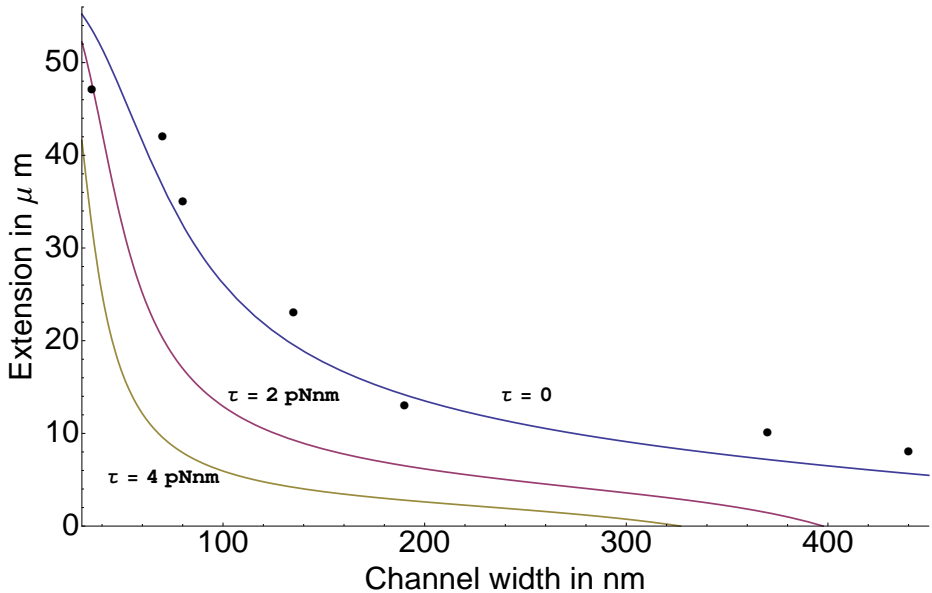}
  \caption{\label{fig:nano}Extension of a 60 $\mu$m DNA chain in square
nanochannels with
widths from 30-450 nm, without torque and with
torques of $2$ and $4$ pNnm ($\tau=0.5,1$), the dots are experimental data from
Ref.~\cite{Reisner:2004}}
\end{figure}

\begin{figure}[htb]
 \includegraphics[width=\columnwidth]{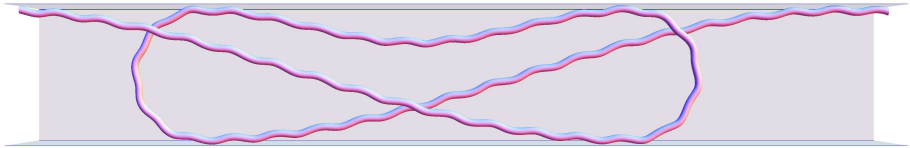}
  \caption{\label{fig:hairpin}Two hairpins forming a planar loop of writhe $\simeq 1$}
\end{figure}

With increasing channel size hairpin formation starts to influence chain
extension~\cite{Odijk:2006} on top of local fluctuations. The free energy of an isolated
hairpin in the mechanical limit, $\bar{F}_{mc}$ is in the torsionless case that of an entropically squeezed mechanical hairpin~\cite{Odijk:2006}. 
The hairpins seen as rare defects in the linear chain have a statistical density that increases exponentially with decreasing free energy of the hairpin. The global existence length is the average distance between hairpins
When a torque is applied to a confined chain approaching the bifurcation point a local minimum will appear starting from an almost closed loop having a writhe close to one. In case the confinement  is replaced by a tension, these loops form the nucleation point of plectonemes, super coiled helices perpendicular to the channel direction. The confinement does not allow for plectoneme formation and so the local minimum is the almost closed loop. These loops  can not grow perpendicular to the channel axis, but can separate in two hairpins  along the axes as shown in Fig.~\ref{fig:hairpin}.
It can be shown~\cite{Emanuel:2012} that when the tangent  of the ground-state
varies on a much larger length scale than the thermal fluctuations, the writhe of
the ground-state and thermal writhe can be treated separately, the total writhe
being their sum. Conceptually the thermal writhe can be calculated from
Fuller's equation with as reference curve the writhing ground-state. As long as
this writhing ground-state is straight on the thermal fluctuations length scale
the writhes separate.
Since the confinement free energy~\eqref{eq:fconfaniso} is only in higher order dependent on the linking number density, torque hardly influences the balance
between entropic confinement and mechanical bending, that determines the hairpin free energy. The writhe of a pair of  successive hairpins on the other hand does decrease the free energy and thereby the extension dramatically. 
In fact one can attribute a writhe of one half to a hairpin since two successive hairpins have this configuration as only minimum in their relative orientation, at least for a circular channel. Some minor contributions of other minima, reflecting the symmetry of the channel cross-section, might be present, but we will neglect them for the rest of the paper.
Assuming the channel to be narrow enough that we can treat the hairpins as a dilute gas of defects we can follow the reasoning of Ref.~\cite{Odijk:2006}. The free energy of a hairpin in the mechanical limit  decreases by the work done by the torque:
$\bar{F}_{mc}(\tau)\simeq \bar{F}_{mc}(0)-\pi \tau$. Or in terms of the global persistence length:
\begin{align}
g(\tau)=g(0)e^{-\pi\tau} 
\end{align}
As example Figure~\ref{fig:nano} shows the effect of moderate $2$ pNnm
($\tau=0.5$) and $4$ pNnm ($\tau=1$) torques on the elongation of a $60$ $\mu$m
DNA chain in a square
channel with widths ranging from $40-100\quad\mu$m, following the theoretical
treatment of Ref.~\cite{Odijk:2006} and in comparison with experiments
from
Reisner
et
al.~\cite{Reisner:2004} on DNA in nanochannels in the absence of torsion. The
shortening by thermal fluctuations was calculated
using equations~\eqref{eq:shortaniso},\eqref{eq:tau} and~\eqref{eq:hard}. As
material
parameters were used $\pb=57$ nm (DNA with dye as in
Ref.~\cite{Reisner:2004})
and
$\pc=100$ nm.
Applying a moderate torque induces its collapse mainly due to
the lowering of the hairpin free energy. Only above $350$ nm does the twist
instability of confinement become important. A moderate torque of $2$
pNnm reduces the
elongation to less than half the value it had in the torqueless
case for channels larger than $90$ nm.

\section{Summary and conclusion}
As is shown in this paper, torsion can have a large impact on the
confinement of semiflexible polymers. For nanosize confinement the
thermal writhe caused by an applied torsion is damped by the tightest
confinement direction. The apparent softening of the chain is similar to the
softening of a semiflexible polymer under torsion~\cite{Moroz:1998} with the
appearance of a new deflection length. The instability that appears in the free
energy landscape sets new limits to the magnitude of the torque the molecule
can propagate depending on the confinement of its environment. It is important
that the strength of the confinement in one direction already suppresses the
thermal writhe and thus increases the torque that propagates.

An important tool in the investigation of biopolymers and their interaction
with proteins is the possibility to spread out the molecule in
nanochannels of
diameter~\cite{Reisner:2004,Jo:2007,Zwolak:2008,reisner2012dna} comparable to
the molecules persistence length. The theory presented here shows that torques
that
for example RNA-polymerase can induce~\cite{Harada:2001} (up to $5$ pN nm) do
increase the demands on channel size considerably. 

\section*{acknowledgments}
We thank Gerhard Gompper, Helmut Schiessel and the referee for many useful remarks and
suggestions for improving the paper and Theo Odijk for a wealth of ideas and
insights. 

\bibliographystyle{plain}
\bibliography{confined}
\end{document}